\documentclass{jpsj-suppl}
\usepackage{txfonts} %Please comment out this line unless the txfonts
\usepackage{graphicx}

\title{Recent Advances in Shell Evolution with Shell-Model Calculations}

\author{Yutaka \textsc{Utsuno}$^{1,2}$, Takaharu \textsc{Otsuka}$^{3,2,4}$, 
Yusuke \textsc{Tsunoda}$^{3}$, Noritaka \textsc{Shimizu}$^{2}$, 
Michio \textsc{Honma}$^{5}$, Tomoaki \textsc{Togashi}$^{2}$, and 
Takahiro \textsc{Mizusaki}$^{6}$}

\inst{$^{1}$Advanced Science Research Center, Japan Atomic Energy Agency,
Tokai, Ibaraki 319-1195, Japan \\
$^{2}$Center for Nuclear Study, University of Tokyo, Hongo Tokyo
 113-0033, Japan \\
$^{3}$Department of Physics, University of Tokyo, Hongo,
 Tokyo 113-0033, Japan \\
$^{4}$National Superconducting Cyclotron Laboratory, Michigan
 State University, East Lansing, MI 48824, USA \\
$^{5}$Center for Mathematical Sciences, University of Aizu, 
Ikki-machi, Aizu-Wakamatsu, Fukushima 965-8580, Japan \\
$^{6}$Institute for Natural Sciences, Senshu University, 
Tokyo 101-8425, Japan
}

\email{utsuno.yutaka@jaea.go.jp}

\recdate{}

\abst{Shell evolution in exotic nuclei is investigated 
with large-scale shell-model calculations. 
After presenting that the central and tensor forces produce 
distinctive ways of shell evolution, we show several 
recent results: (i) evolution of single-particle-like levels 
in antimony and cupper isotopes, 
(ii) shape coexistence in nickel isotopes understood in terms of 
configuration-dependent shell structure, 
and (iii) prediction of the evolution of the recently established 
$N=34$ magic number towards smaller proton numbers. 
In any case, large-scale shell-model calculations play indispensable roles 
in describing the interplay between single-particle character and 
correlation. 
}

\kword{exotic nuclei, shell evolution, large-scale shell-model calculation, 
magic number}

\begin{document}
\maketitle

\section{Introduction---shell evolution due to the nuclear force}

The evolution of shell structure in nuclei, often called shell
evolution, attracts much interest 
in recent investigations in the physics of exotic nuclei 
\cite{gade_rev, sorlin_rev, otsuka_rev}. 
Shell evolution can be accessed through the systematics of 
an isotope (isotone) chain, where the proton (neutron) Fermi
surface is kept unchanged. 
Such systematic data along long isotope chains 
are being taken due to recent advances in 
experiment using radioactive-ion (RI) beams, and they disclose  
various phenomena that indicate quite sharp shell evolution 
compared to that of the conventional one-body potential. 
The most prominent case is 
the breakdown of the traditional magic numbers including 
the disappearance of the $N=20$ magic number 
known as ``island of inversion'' \cite{island} 
and the emergence of an $N=16$ magic number \cite{n16nature}. 

While shell evolution is considered as a universal
property of nuclei, its complete behavior over the nuclear chart 
is still ambiguous. 
This is partly because extracting the evolution of single-particle
levels from experiment is not straightforward even in the vicinity of the magic 
numbers. For instance, having the proton number 51, 
antimony isotopes are sometimes considered to have nearly pure 
proton single-particle states in the low-lying energy region
\cite{schiffer}. 
On the other hand, it has been pointed out \cite{sorlin_rev} that 
those single-particle-like states are strongly mixed with the states
involving core excitation (core-coupled states)
because typical $2^+_1$ and $3^-_1$ 
excitation energies in tin isotopes are $\sim 1$ and $\sim 2$ MeV, 
respectively, 
being close to single-particle-energy spacings. 
This possible mixing with the core-coupled state makes it difficult 
to quantify the single-particle levels 
without resorting to large-scale nuclear-structure calculations 
which suitably treat many-body correlations including 
collective $2^+$ and $3^-$ states. 

\begin{figure}[t]
\begin{center}
\includegraphics[width=11.0cm,clip]{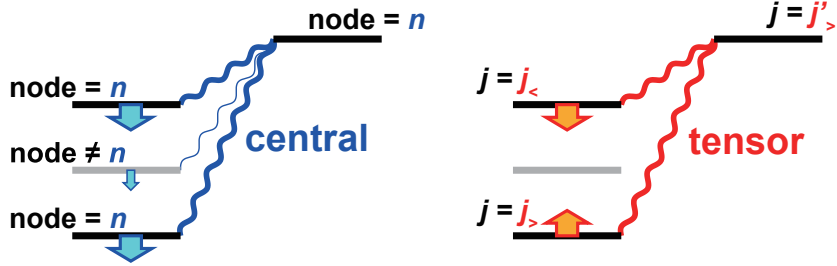}
\caption{Schematic illustration of shell evolution caused by the central
 and tensor proton-neutron forces. }
\label{fig:force}
\end{center}
\end{figure}

The phenomena of shell evolution call for 
a new mechanism of the formation of shell
structure beyond the conventional one-body potential picture.
The role of the nuclear force has recently received much attention. 
If two nucleons in the orbits $i$ and $j$ receive mean
attraction whose strength is larger than other pairs, 
then the $i$ orbit becomes bound more deeply than the others 
when the orbit $j$ is occupied. 
Here, mean attraction or repulsion stands for angular-momentum averaged 
two-body matrix elements called the monopole interaction 
\cite{monopole}. 
In Fig.~\ref{fig:force}, two major sources that cause the shell evolution 
in the proton-neutron channel---the central force and the tensor
force---are presented together with their effects on the shell structure. 
Those forces induce very different shell evolutions. 
The central force, whose effect has been known for a long time 
(see \cite{heyde91} for instance), gives rise to a stronger attraction 
between the nucleons in the orbits of the same node. 
Its dependence on spin direction is rather weak. 
The tensor force, on the other hand, produces the monopole matrix 
elements that are opposite in sign between the spin-orbit partners 
$j_>=l+1/2$ and $j_<=l-1/2$ due to the identity 
$(2j_>+1)v_{j_>,j'}^{T}+(2j_<+1)v_{j_<,j'}^{T}=0$
derived by Otsuka {\it et al.} in 2005 \cite{tensor}, 
where $v_{i,j}^T$ denotes the monopole matrix element between two
nucleons in the $i$ and $j$ orbits with isospin coupling $T$. 
As a result, the tensor force drastically changes the spin-orbit splitting. 
From the viewpoint of the Nilsson model, the central force and the tensor force 
are responsible for the evolution of the 
\mbox{\boldmath $l$}$^2$ and the \mbox{\boldmath $l\cdot s$}
terms, respectively. 

The above-mentioned concept has been quantitatively implemented in  
the ``monopole-based universal interaction'' ($V_{\rm MU}$) which 
aims at a universal description of shell evolution due to the
nuclear force \cite{vmu}. 
The original $V_{\rm MU}$ \cite{vmu} consists of the central force and the tensor force. 
Its tensor force, the $\pi + \rho$ meson exchange potential, is
supported microscopically by the ``renormalization persistency''
\cite{vmu, ren} 
which states that the tensor matrix elements are almost unchanged by 
the renormalization of the nuclear force. 
The effective central force includes most of the renormalization effect,
and should be quite complicated in principle. 
However, an analysis of semi-empirical effective interactions 
for the $sd$ and $pf$ shells shows that the monopole matrix elements 
for the central force seem to be rather simple. 
This result motivates us to phenomenologically describe 
the central force in a simple functional form. 
In the $V_{\rm MU}$, 
a one-range Gaussian central force is employed. 
This choice is successful in well approximating 
the central part of the monopole matrix elements of the 
SDPF-M \cite{sdpfm} interaction for the $sd$ shell 
and GXPF1A \cite{gxpf1a} interaction for the $pf$-shell 
simultaneously. 
The $V_{\rm MU}$ interaction provides reasonable 
shell evolutions in various regions of recent interest \cite{vmu} 
as far as single-particle-like levels are concerned. 

The concept of shell evolution should be verified 
within reliable many-body calculations 
because most of the actual states are highly correlated. 
Our group has been approaching this issue on the basis 
of large-scale shell-model calculations. 
In this paper, we report on very recent progress in 
large-scale shell-model calculations for exotic nuclei 
from the viewpoint of shell evolution. 
Methodological and computational aspects of the large-scale shell-model 
study are covered in this conference by Shimizu \cite{shimizu}. 

\section{Evolution of single-particle-like states}
\label{sec:spe}

In this section, single-particle-like states in antimony 
and copper isotopes are examined. 
Since antimony ($Z=51$) is located one-major-shell above copper ($Z=29$), 
one expects similarity in shell evolution between the two isotope
chains. 
It is shown that correlation cannot be neglected 
even for those single-particle-like states. 

\begin{figure}[t]
\begin{center}
\includegraphics[width=9.0cm,clip]{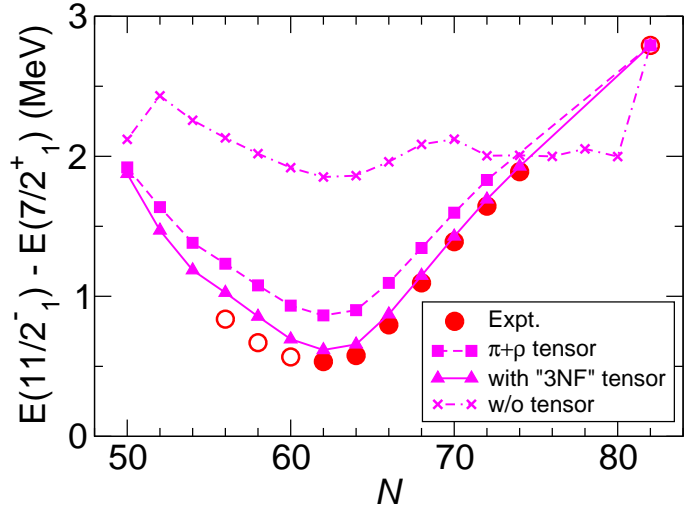}
\caption{11/2$^-_1$ levels in antimony isotopes relative to 
$7/2^+_1$ levels. Experimental data are compared to 
three shell-model calculations using different 
strengths of the tensor force. The 3NF contribution is 
phenomenologically introduced by multiplying the $T=0$ 
part of the $\pi+\rho$ tensor force by 1.3 
according to a recent suggestion \cite{kohno}. 
}
\label{fig:sb}
\end{center}
\end{figure}

\subsection{Antimony isotopes}
\label{sec:sb}

In antimony isotopes, the $7/2^+_1$ and the $11/2^-_1$ levels are observed 
along a long isotope chain. Those levels are dominated by 
proton $0g_{7/2}$ and $0h_{11/2}$ single-particle wave functions,
respectively, thus providing useful information on shell evolution. 
Since $0g_{7/2}$ and $0h_{11/2}$ are $j_<$ and $j_>$ orbits, respectively, 
the evolution of the energy spacing between 
those levels is sensitive to the evolution 
of spin-orbit splitting 
as pointed out by Schiffer {\it et al.} \cite{schiffer}. 
While this single-particle evolution is excellently accounted for 
by introducing the tensor force \cite{tensor}, 
significant mixing with core-coupled wave functions 
could make the situation rather complicated 
\cite{sorlin_rev}. 

We perform large-scale shell-model calculations for antimony isotopes 
from $N=50$ to $N=82$ in order to confirm the dominance of the tensor 
force in the evolution of energy levels within a 
reliable many-body framework. 
The valence shell taken in this calculation consists of 
the $0g_{7/2}$, $1d_{5/2}$, $1d_{3/2}$, $0h_{11/2}$ and $2s_{1/2}$ 
orbits. The neutron-neutron effective interaction is responsible 
for constructing good wave functions of tin cores. 
We use a semi-empirical interaction named SNBG3 \cite{snbg1}
for this part, with 
which low-lying energy levels of tin isotopes are described 
within a few hundred keV. 
The proton-neutron effective interaction, on the other hand, 
controls shell evolution. We take 
the density-dependent version of the $V_{\rm MU}$ interaction 
used for the shell-model calculation in the $sd$-$pf$ shell 
\cite{sdpfmu}, 
and apply to it a scaling factor 0.84 for the central 
part to reproduce good proton separation energies. 
The proton single-particle energies are determined to fit 
the energy levels in $^{133}$Sb. 

In  Fig.~\ref{fig:sb}, we compare 
the $11/2^-_1$ energy levels relative to $7/2^+_1$
among three different tensor forces. 
In each case, the other two-body 
forces are the same, and the bare proton single-particle 
energies are readjusted to fit the energy levels in $^{133}$Sb.
One sees a clear effect of the tensor force. 
When the tensor force is omitted, the $11/2^-_1$ levels 
are located almost constantly $\sim 2$~MeV above the 
$7/2^+_1$ levels. 
In this case, dominated by core-coupled states, 
the $11/2^-_1$ levels contain little proton 
$0h_{11/2}$ single-particle component. 
On the other hand, shell-model calculations with the $\pi+\rho$ 
tensor force lead to reasonable agreement with experiment. 
When a slightly stronger tensor force is used to include 
the effect of three-nucleon force as suggested in Ref.~\cite{kohno}, 
agreement with experiment becomes almost perfect. 

Although the evolution of the $11/2^-_1$-$7/2^+_1$ spacing is strongly 
influenced by the tensor-force-driven shell evolution, 
our calculation shows that those energy levels cannot be 
regarded as single-particle states. The calculated proton 
spectroscopic factors for those levels are 0.5-0.7 
in the mid-shell region ($N\sim 64$), which are much smaller 
than unity. The increase in correlation towards the mid-shell 
is strongly supported by the evolution of magnetic moments. 
The magnetic moments in the $7/2^+_1$ states measured from $N=70$ to 
$N=82$ show a gradual deviation from the Schmidt value 
with decreasing neutron number from $N=82$. 
The present calculation is successful in 
reproducing this change in parallel with the decrease in 
spectroscopic factors. It is interesting to see that 
in spite of significant non-single-particle contribution 
the evolution of the $11/2^-_1$-$7/2^+_1$ spacing is well described 
with the single-particle evolution due to the tensor force 
\cite{tensor}. The calculation shows that 
this takes place because of cancellation in energy 
between neutron-neutron correlation and proton-neutron correlation.

\subsection{Copper isotopes}
\label{sec:cu}

A sharp shell evolution between the proton $1p_{3/2}$ and $0f_{5/2}$ orbits 
is predicted to occur in going from $N=40$ to $N=50$, where 
the neutron $0g_{9/2}$ orbit is filled. 
This is due to a cooperative effect of the central force and the tensor 
force: both favor the $\pi 0f_{5/2}$-$\nu 0g_{9/2}$ monopole 
matrix elements more than that of  $\pi 1p_{3/2}$-$\nu 0g_{9/2}$
(see Fig.~\ref{fig:force}). As a result, the proton $0f_{5/2}$ 
orbit sharply drops down compared to the $1p_{3/2}$ orbit 
with increasing neutron number as shown in Ref.~\cite{vmu}. 
This effect is found in the copper isotope chain. The ground states 
of copper isotopes are known to have $3/2^-$ for $N\le 44$, 
whereas the ground state of the $N=46$ isotope has recently been 
measured to be $5/2^-$ \cite{flanagan}. 

\begin{figure}[t]
\begin{center}
\includegraphics[width=8.0cm,clip]{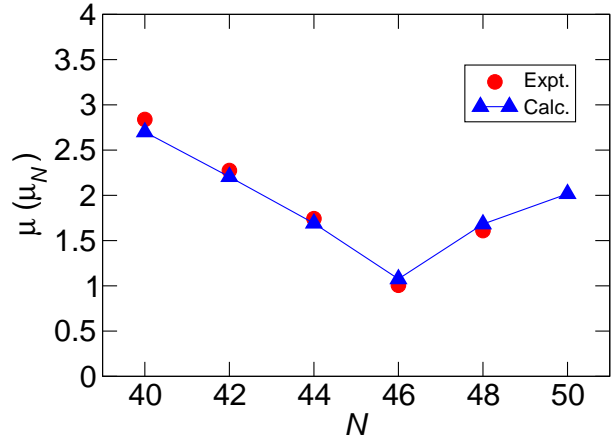}
\caption{Magnetic moments in the ground states of Cu isotopes 
($3/2^-$ for $40\le N \le 44$, and $5/2^-$ for  $46\le N \le 50$)
compared between experiment and the MCSM calculation. 
Experimental data are taken from Refs.~\cite{flanagan, koster}. 
In the MCSM calculation, spin $g$ factors are quenched by 0.7, and 
isovector orbital $g$ factors are added to 
the free-nucleon $g$ factors by  
$\delta g_l({\rm IV})=(\delta g_l(p)-\delta g_l(n))/2=0.1$. 
}
\label{fig:cu}
\end{center}
\end{figure}

As discussed for the antimony isotopes in Sect.~\ref{sec:sb}, 
magnetic moments provide crucial 
information on to what extent the single-particle state dominates 
the many-body state. For copper isotopes, the measured magnetic 
moments have been compared to the single-particle estimates and 
shell-model calculations in the 
valence shell consisting of the $1p_{3/2}$, $0f_{5/2}$, 
$1p_{1/2}$ and $0g_{9/2}$ orbits in Ref.~\cite{flanagan}. 
The measured moments start to depart from the Schmidt value 
with increasing neutron number from $N=40$. 
Although this tendency is described with the shell-model calculation, 
a certain deviation from experiment still remains 
especially in $^{73}$Cu. This deviation has disappeared in a later 
shell-model calculation that activates the proton $0f_{7/2}$ orbit
to allow proton excitation 
across the $Z=28$ gap \cite{cu_sieja}. This indicates that very 
large-scale shell-model calculations are required to describe 
the structure of copper isotopes because the magic number 28 is more 
fragile than the ones larger than 28. 
We carry out such large-scale 
shell-model calculations in the full $pf$ shell plus 
the $0g_{9/2}$ and $1d_{5/2}$ orbits using the advanced Monte Carlo 
shell-model (MCSM) calculation \cite{rmcsm}. 
The effective interaction adopted here is the same as the one 
used in the study of nickel isotopes (see Sect.~\ref{sec:ni} and 
Ref.~\cite{type2}). As shown in Fig.~\ref{fig:cu}, the magnetic 
moments calculated with MCSM are in excellent agreement with the 
experimental data. 
In this way, we achieve a unified description of nickel and copper 
isotopes. 

While the structure of neutron-rich copper isotopes is well described 
with large-scale shell-model calculations of Ref.~\cite{cu_sieja} and 
the present one, 
it is still ambiguous how much the $0f_{5/2}$ orbit lowers 
relative to $1p_{3/2}$ in going from $N=40$ to $50$. 
Rather large changes ($4.5$-$5$~MeV) are predicted in Ref.~\cite{cu_sieja} 
and the JUN45 interaction \cite{jun45}. 
On the other hand, $V_{\rm MU}$ and the present calculation lead to 
smaller changes ($3$-$3.5$~MeV). 
The energy levels in $^{79}$Cu, the $N=50$ isotope, will provide 
crucial information to quantitatively evaluate the evolution 
of the $0f_{5/2}$ and $1p_{3/2}$ orbits.

\section{Shape coexistence in nickel isotopes and 
Type II shell evolution}
\label{sec:ni}

The structure of neutron-rich $N\sim 40$ nuclei 
has recently been very extensively investigated in RI-beam 
facilities over the world. One of the interesting issues 
in this region is a sudden change of nuclear deformation 
in going down from nickel to smaller proton numbers. 
This situation is analogous to the so-called island of inversion 
around $^{32}$Mg \cite{island}. 

Concerning shell evolution, 
it is argued in 
Ref.~\cite{lnps} that the $N=40$ shell gap between the $pf$ shell 
and $0g_{9/2}$ narrows in going to lower proton numbers 
in a similar way to the evolution of the $N=20$ shell gap 
realized with the SDPF-M interaction \cite{sdpfm}. 
In this usual sense of shell evolution, 
nickel isotopes have shell gaps large 
enough to stabilize spherical shapes in their ground states. 
In contrast, we have proposed an unknown manifestation 
of shell evolution in Ref.~\cite{type2}: 
the shell gap is not necessarily fixed in a designated nucleus 
but can evolve depending on configuration. Here we consider 
how the shell evolution within a nucleus influences shape coexistence 
in nickel isotopes. The spherical configurations 
in nickel isotopes are the filling configurations, and 
produce relatively large $Z=28$ and $N=40$ shell gaps. 
On the other hand, the deformed configurations are associated with 
particle-hole excitation across the $N=40$ shell gap 
(e.g., $4p$-$4h$ configuration for the prolate state). 
Thus, the deformed configurations have more neutron particles in the 
$0g_{9/2}$ orbit and neutron holes in the $0f_{5/2}$ and 
$1p_{1/2}$ orbits than the spherical
configurations. The occupation in the $j_>$ orbit and absence in  
the $j_<$ orbit work coherently to reduce the spin-orbit 
splitting of proton orbits due to the tensor force. 
Consequently, protons are more easily 
excited into the upper $pf$ shell, and deformation is promoted. 
This is rather different from the conventional one-body potential 
picture such as the Nilsson model in which the shell structure is 
constant within a fixed proton-neutron number. 
The shell evolution which occurs within a nucleus depending on
configuration is called Type II shell evolution 
to distinguish from the usual (Type I) shell evolution 
which is based on the filling 
configurations. 

In Ref.~\cite{type2}, a unified description of nickel isotopes 
from $N=28$ to $N=50$ has been given 
with the MCSM calculations in the full $pf$ shell 
plus the $0g_{9/2}$ and $1d_{5/2}$ orbits. We have developed a new 
method to visualize shape fluctuation in each strongly correlated 
many-body wave function by utilizing the intrinsic shapes of MCSM 
basis states \cite{type2}. With this method, we have clearly shown that 
spherical-oblate-prolate shape coexistence occurs in $^{68}$Ni. 
Type II shell evolution is crucial in stabilizing deformation. 
Similar shape coexistence is predicted for surrounding nuclei, 
which is of great interest to confirm in forthcoming experiments. 

\section{Evolution of the new $N=34$ magic number}

It has been predicted in 2001 that 
a new $N=34$ magic number 
appears around calcium isotopes 
as a large shell gap between $1p_{1/2}$ and $0f_{5/2}$
\cite{magic}. 
The mechanism of its appearance is the same as that of the $N=16$ 
magic number for the $sd$ shell \cite{sdpfm,ozawa}: 
strong attraction between a proton in a $j_>$ orbit 
($0d_{5/2}$ or $0f_{7/2}$) and a neutron in a $j_<$ orbit 
($0d_{3/2}$ or $0f_{5/2}$) produces a large shell gap below 
the neutron $j_<$ orbit 
when no protons occupy the $j_>$ orbit (i.e., oxygen or calcium). 
As for the $pf$ shell, the 
$\nu 0f_{5/2}$ orbit is known to be located between 
$\nu 1p_{3/2}$ and $\nu 1p_{1/2}$ 
in nickel isotopes. When protons are removed from $0f_{7/2}$, 
the $\nu 0f_{5/2}$ orbit is shifted up above the $\nu 1p_{1/2}$ 
orbit. The $N=32$ shell gap, the one between 
$\nu 1p_{3/2}$ and $\nu 1p_{1/2}$, thus appears first. Several evidences 
for the $N=32$ magic number have been found in chromium, titanium, and 
calcium isotopes \cite{cr56, ti54, ca52}. 
On the other hand, the $N=34$ shell gap is not large enough 
to make a closed-shell structure in chromium and titanium isotopes 
\cite{cr56,ti56}. 
The occurrence of the $N=34$ magic number has been observed 
quite recently \cite{ca54} from the $2^+_1$ level in $^{54}$Ca
\cite{ca54} which is much higher than the ones in non-magic nuclei 
such as $^{42-46,50}$Ca. Those experimental results 
indicate that the $N=34$ magic number is quite localized 
because of a sharp evolution of the $\nu 0f_{5/2}$ orbit as a function 
of the proton number. 

Although the strength of the $N=34$ shell gap cannot be directly 
measured, it is estimated from the comparison of energy levels 
between experiment and shell-model calculations. We take the 
GXPF1B \cite{gxpf1b} and GXPF1Br \cite{ca54} interactions for  
the full $pf$-shell calculations. The GXPF1B interaction 
gives a good description 
of neutron-rich calcium isotopes systematically, but 
its $2^+_1$ level in $^{54}$Ca is 0.6~MeV too high. 
To remedy this small discrepancy in a simple way, 
the GXPF1Br interaction is proposed \cite{ca54} by minimally 
modifying the GXPF1B interaction, 
$\delta v^{T=1}_{1p_{3/2},0f_{5/2}}=-0.15$~MeV, 
in order to reproduce the 
$2^+_1$ level in $^{54}$Ca. The resulting $N=34$ shell gap for 
the GXPF1Br interaction is 2.66 MeV. The GXPF1Br interaction 
improves not only the $2^+_1$ level in $^{54}$Ca but also 
some other energy levels in calcium and scandium isotopes 
which are dominated by the configurations involving $0f_{5/2}$. 
Thus, the $N=34$ shell gap in calcium isotopes is estimated 
to be $\sim 2.6$~MeV, which is close to the $N=32$ shell gap. 

Once a large $N=34$ shell gap in calcium isotopes is established, 
it is a natural question to ask how this magic structure evolves 
and how the evolution affects observables 
in the more exotic region, i.e., isotopes with less proton numbers. 
Since the proton valence shell moves to the $sd$ shell, 
the proton-neutron monopole matrix elements relevant to 
the evolution of the $N=34$ shell gap are 
$\pi 0d_{3/2}$-$\nu 1p_{1/2}$ vs. $\pi 0d_{3/2}$-$\nu 0f_{5/2}$ and 
$\pi 1s_{1/2}$-$\nu 1p_{1/2}$ vs. $\pi 1s_{1/2}$-$\nu 0f_{5/2}$. 
Concerning the matrix elements associated with $\pi 0d_{3/2}$, 
the $\nu 0f_{5/2}$ orbit is more attractive 
than the  $\nu 1p_{1/2}$ orbit for the central force, but is 
more repulsive for the tensor force.
The matrix elements regarding $\pi 1s_{1/2}$ orbit favor 
$\nu 1p_{1/2}$ more than $\nu 0f_{5/2}$ in the central channel, and 
vanish in the tensor channel. 
Thus, the evolution of the $N=34$ magic number towards smaller 
proton numbers is difficult to predict only from this qualitative 
discussion. 

Those monopole matrix elements are quantitatively estimated 
on the basis of the SDPF-MU interaction \cite{sdpfmu}, 
whose cross-shell part consists of the $V_{\rm MU}$ interaction. 
Although its cross-shell monopole matrix elements
are quite reasonable on the whole
in view of its descriptive power \cite{sdpfmu}, there is room for 
improvement on the central matrix element between $1s$ and $1p$ 
because of some discrepancy in the evolution of the $\pi 1s_{1/2}$ 
hole levels in potassium isotopes beyond $N=28$ \cite{potasium}. 
The binding energies and energy levels in neutron-rich 
potassium isotopes strongly constrain the amplitude of the 
central monopole matrix elements between proton $0d$/$1s$ and 
neutron $0f$/$1p$ orbits. 
Here we shift 
the $\pi 1s$-$\nu 1p$ central monopole matrix elements by $+0.3$~MeV 
so as to achieve good agreement with experimental energies in 
potassium isotopes. In addition to this change, the SDPF-MU interaction 
is modified by replacing its $pf$-shell part with 
the GXPF1Br interaction. 

\begin{figure}[t]
\begin{center}
\includegraphics[width=13.0cm,clip]{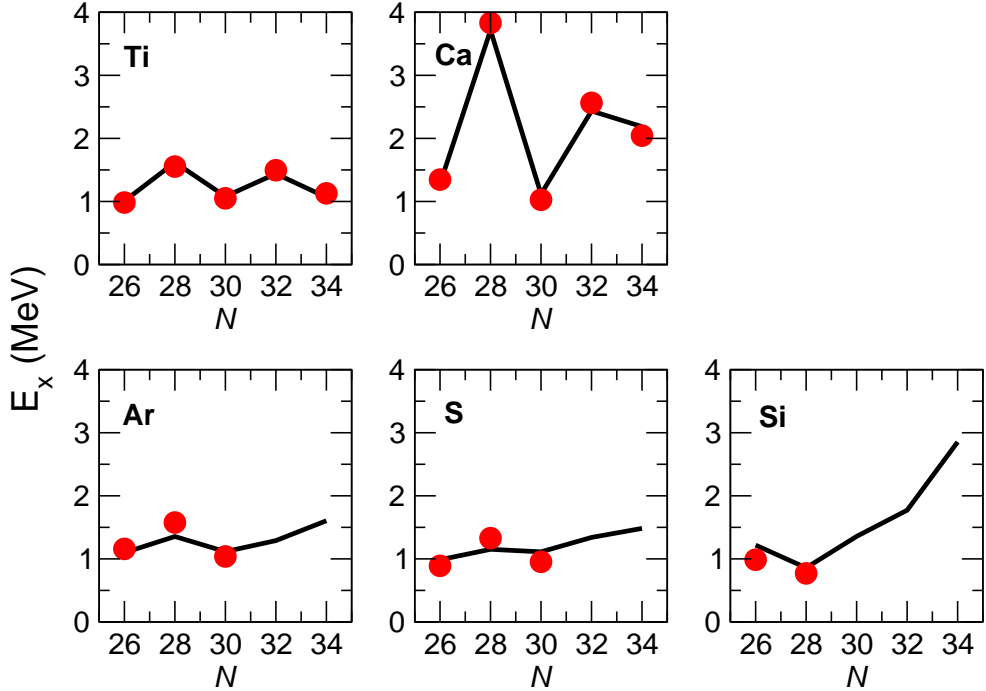}
\caption{$2^+_1$ energy levels in neutron-rich 
titanium, calcium, argon, sulfur, and silicon 
isotopes compared between experiment 
(circles) and theory (lines). 
}
\label{fig:level}
\end{center}
\end{figure}

The resulting modified SDPF-MU interaction predicts the enhancement of 
the $N=34$ shell gap towards lower proton numbers: 
it ends up with 3.9 MeV at $^{48}$Si. 
Although the evolution of the $N=34$ shell gap has not been observed
for $N=34$ isotones, 
the shell gap between $\nu 1p_{1/2}$ and $\nu 0f_{5/2}$ appears to 
somewhat enhance in going from calcium to silicon 
in terms of the distribution of spectroscopic factors in $N=20$ isotones 
\cite{si35}. 
According to shell evolution due to the monopole interaction, 
if the $N=34$ shell gap enlarges for the $N=20$ core to some extent, 
it enlarges for the $N=34$ core to the same degree. 
Hence, it is likely that the $N=34$ shell gap enlarges towards silicon 
isotopes. 
The effect of the enhanced $N=34$ shell gap 
can be detected experimentally. In Fig.~\ref{fig:level}, 
the $2^+_1$ energy levels 
in neutron-rich titanium, calcium, argon, sulfur and silicon isotopes
are compared between the experimental data 
and the shell-model calculations using the modified SDPF-MU 
interaction. While the $2^+_1$ energy in titanium isotopes 
decreases in going from $N=32$ to $N=34$, the ones in argon, 
sulfur and silicon isotopes increase due to their 
large $N=34$ shell gaps. In particular, $^{48}$Si is predicted to 
be a new doubly magic nucleus, having a $2^+_1$ level located at 
$\sim 3$~MeV. It is of great interest to verify this prediction 
in current and future RI-beam facilities. 

\section{Conclusion}

We have performed large-scale shell-model calculations for exotic nuclei 
ranging from light to medium-heavy regions 
to investigate shell evolution. Currently, the central and the tensor forces 
are regarded as two major sources that cause shell evolution 
in the proton-neutron channel. Since those two forces cause rather different
evolutions in character, it is possible to extract the
tensor-force-driven shell evolution from experimental data 
concerning exotic nuclei. 
Large-scale shell-model calculations are of great help for this 
purpose. We have calculated single-particle-like levels in antimony 
and copper isotopes systematically. 
Although those levels suffer much correlation, their evolutions are 
strongly dominated by shell evolution. As a result, 
the current understanding of shell evolution implemented in $V_{\rm MU}$ 
is confirmed. We point out that shape coexistence in nickel isotopes 
is stabilized through a new manifestation of shell evolution, 
Type II shell evolution. We finally predict that the new $N=34$ 
magic number recently found in $^{54}$Ca 
persists with decreasing proton numbers. 
The $N=34$ shell gap is predicted to enlarge there, and 
measuring the $2^+_1$ energies 
in very neutron-rich argon, sulfur, and silicon isotopes 
will provide crucial information to probe the predicted evolution.

\section*{Acknowledgments}
This work was supported in part by JSPS KAKENHI Grant Numbers 
21740204 and 23244049 and by the 
HPCI System Research Project (hp120284, hp130024 and hp140210). 
This work is a part of the CNS-RIKEN joint research project on
 large-scale nuclear-structure calculations.

\end{document}